\documentclass[aps,prl,twocolumn,groupedaddress,showpacs]{revtex4-1}

\usepackage{graphicx}
\usepackage{amsmath,gensymb}
\usepackage{subfigure}
\usepackage{amssymb}
\usepackage{amsbsy}
\usepackage{tabularx}
\usepackage{epstopdf}
\usepackage{epsfig}
\usepackage{color}
\usepackage{bm}
\usepackage{setspace}

\newcommand{\thetain}{\theta_{\textrm{in}}}
\newcommand{\thetaout}{\theta_{\textrm{out}}}
\newcommand{\avgthout}{\left<\theta_{\textrm{out}}\right>}
\newcommand{\thetars}{\theta_{\textrm{rs}}}
\definecolor{amber}{rgb}{1.0, 0.49, 0.0}

\begin{document}

\widetext

\title{Microalgae scatter off solid surfaces by hydrodynamic and contact forces}
\author{Matteo Contino$^1$, Enkeleida Lushi$^2$, Idan Tuval$^3$, Vasily Kantsler$^1$ and Marco Polin$^1$}
\email{M.Polin@warwick.ac.uk}

\affiliation{$^1$Physics Department, University of Warwick, Gibbet Hill Road, Coventry CV4 7AL, United Kingdom\\
$^2$School of Engineering, Brown University, Rhode Island 02912, USA\\
$^3$Mediterranean Institute for Advanced Studies (CSIC-UIB), E-07190 Esporles, Spain}

\begin{abstract}
Interactions between microorganisms and solid boundaries play an important role in biological processes, like egg fertilisation,  biofilm formation and soil colonisation, where microswimmers move within a structured environment. Despite recent efforts to understand their origin, it is not clear whether these interactions can be understood as fundamentally of hydrodynamic origin or hinging on the swimmer's direct contact with the obstacle.
Using a combination of experiments and simulations, here we study in detail the interaction of the biflagellate green alga \textit{Chlamydomonas reinhardtii}, widely used as a model puller microorganism, with convex obstacles, a geometry ideally suited to highlight the different roles of steric and hydrodynamic effects.  Our results reveal that both kinds of forces are crucial for the correct description of the interaction of this class of flagellated microorganisms with boundaries.

\end{abstract}
\maketitle
Microorganismal motility is often confined by solid objects. From biofilm formation within soil's porous structure \cite{durham12}, to protistan parasites navigating through the densely packed blood of the host \cite{heddergott12}, and mammalian ova fertilisation \cite{denissenko12}, solid  boundaries alter both motion and spatial distribution of microorganisms \cite{galajda07, kantsler13} in ways that are currently not well understood \cite{berke08, li09}. 
Explaining these interactions can pave the way for the use of extant microorganisms in technological applications ranging from bioremediation  \cite{valentine10, kessler11}, to directed transport and delivery of pharmacological cargo at the microscale \cite{weibel05a}, as well as inform the design of artificial microswimmers \cite{brown14}.
One of the most basic types of interaction is the scattering off a solid plane. Bacteria and other microswimmers with rear-mounted flagella (``pusher''-type) are well known to accumulate spontaneously on planar surfaces \cite{frymier95}, a phenomenon that has been equally well explained by theories based on either purely steric \cite{li09} or hydrodynamic \cite{berke08, ardekani14} interactions. New experiments are finally prising these two effects apart, with results in clear support of the latter \cite{molaei14, sipos15}.
Our knowledge of cell-wall interaction for the other major class of microswimmers, those with front-mounted flagella (``puller''-type), is distinctly less advanced. 
Recent experiments suggest that steric effects dominate the scattering of these flagellates off flat boundaries \cite{kantsler13}. If true in general, this would place the two microswimmer types in clearly separated categories of interaction. However, similarly to the bacterial case \cite{sipos15}, differentiating steric and hydrodynamic effects requires one to move beyond a plane wall.
Here we report the first detailed experimental study of the scattering of a model puller-type microswimmer, the biflagellate alga {\it Chlamydomonas reinhardtii} (CR) \cite{goldstein15}, off a curved surface. Our results, supported also by numerical simulations, show that both hydrodynamic and steric forces are needed to explain the microswimmer's interaction with obstacles. At close contact, lubrication forces alone can lead to long-term entrapment, which is avoided through direct flagellar action following cell spinning. 
\begin{figure}[b!]
\includegraphics[width = 0.95\columnwidth ]{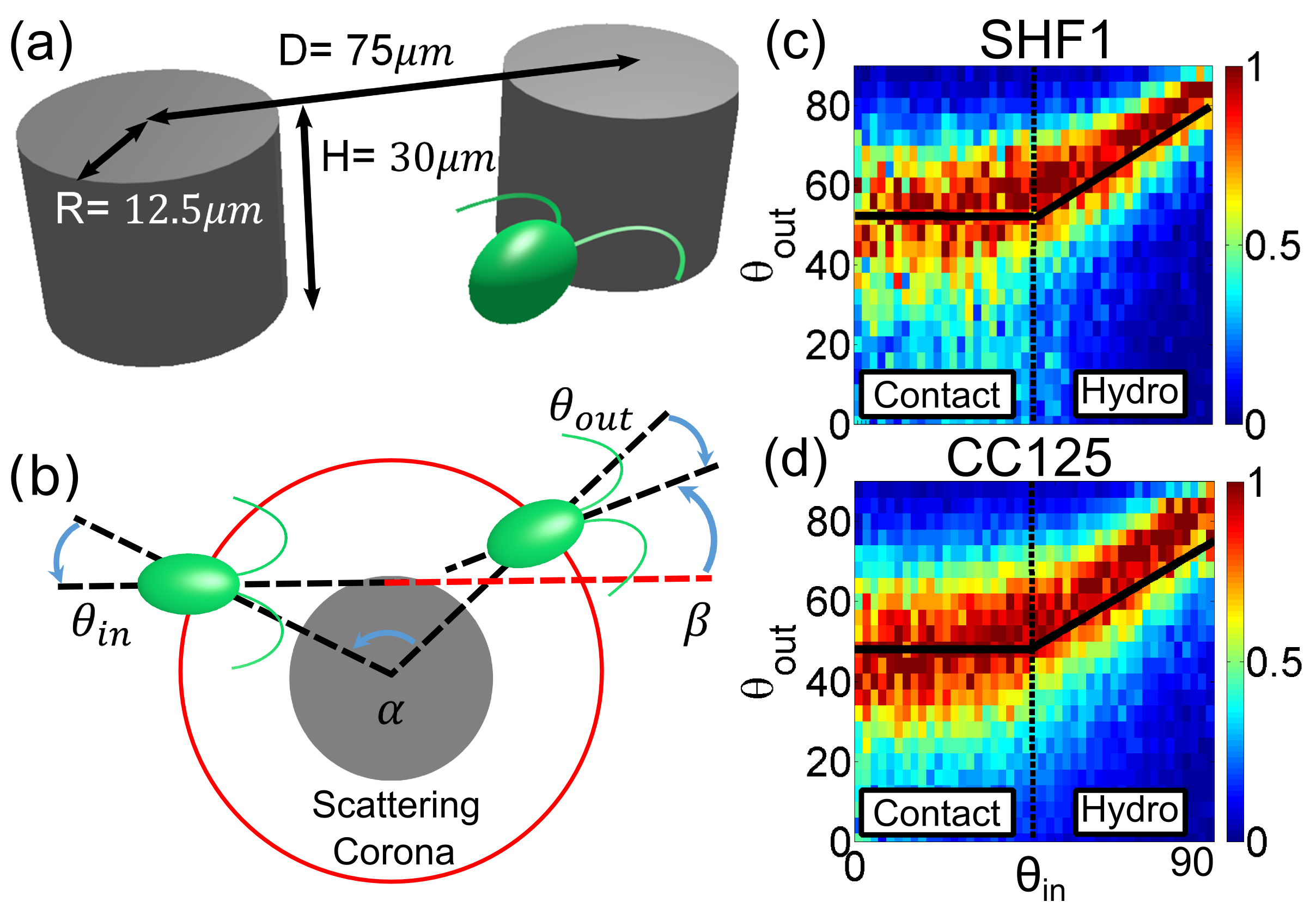}
\caption{Experimental configuration and scattering functions. (a) Schematics of pillars' arrangement. (b) Conventions used for the (signed) scattering angles. (c) Scattering probability $p(\thetaout|\thetain)$ for CC125 and SHF1. Solid lines: fits to Eq.~\ref{eq:thetaout}.}
\label{fig1}
\end{figure}
\begin{table*}[t]
\begin{tabularx}{0.95\textwidth}{l |>{\centering\arraybackslash}X|>{\centering\arraybackslash}X|>{\centering\arraybackslash}X|>{\centering\arraybackslash}X|>{\centering\arraybackslash}X|>{\centering\arraybackslash}X|>{\centering\arraybackslash}X|} 

\multicolumn{1}{c}{     }
&  \multicolumn{1}{c}{ $\theta^*_{\textrm{in}}$}
& \multicolumn{1}{c}{$\theta^*_{\textrm{out}}$} 
& \multicolumn{1}{c}{$m$}
& \multicolumn{1}{c}{$q$}
& \multicolumn{1}{c}{$\textrm{d}_{min}(\theta^*_{\textrm{in}})\,$}
& \multicolumn{1}{c}{$\ell$}
& \multicolumn{1}{c}{$\thetars$}
\\

\cline{2-8}
 CC125 & $44^\circ$ & $(48\pm\,1)^\circ$ & $0.59\pm\,0.01$ & $(22\pm\,0.5)^\circ$ & $10.7\pm\,0.1\,\mu$m & $11.2\pm\,0.2\,\mu$m & $(33.3\pm 0.4)^\circ$ \\
\cline{2-8}
SHF1 & $45^\circ$ & $(50.76\pm\,1)^\circ$ & $0.64\pm\,0.05$  & $(21\pm\,0.4)^\circ$ & $9.0\pm\,0.2\,\mu$m & $9.36\pm\,0.2\,\mu$m & $(50.0\pm\,0.7)^\circ$ \\
\cline{2-8}
\cline{2-8}
CC125$_{\textrm{sim}}$ & $53^\circ$ & $(45\pm1)^\circ $ & $0.78$  & $7.23^{\circ}$ & $1.58a_B$ & $2a_B$ & $--$ \\
\cline{2-8}
SHF1$_{\textrm{sim}}$ & $51.5^\circ$ & $(44.3\pm 1)^\circ$ & $0.83$  & $1.3^{\circ}$ & $1.41a_B$ & $1.8a_B$ & $--$ \\
\cline{2-8}
\end{tabularx}
\caption{Synopsis of experimental and simulation parameters. $(\theta^*_{\textrm{in}}, \theta^*_{\textrm{out}}, m, q)$ are defined in Eq.~\ref{eq:thetaout}; $\textrm{d}_{min}(\theta^*_{\textrm{in}})$: minimal distance of {\it Chlamydomonas} from the pillar surface for $\thetain=\theta^*_{\textrm{in}}$; $\ell$: average flagellar length; $\thetars$: characteristic decay angle for the probability of random scattering.}
\label{tab:values}
\end{table*}
 
CR strains CC125 and SHF1 (short flagella mutant) were grown axenically in Tris-Acetate-Phosphate medium \cite{rochaix88} at $21\degree$C under continuous fluorescent illumination ($100\,\mu$E/m$^2$s, OSRAM Fluora). Cells from exponentially growing cultures at $\sim5\times10^6$cells/ml were harvested and loaded into $30\,\mu$m thick PDMS-based microfluidic channels (Fig.~\ref{fig1}a) previously passivated with a Bovine Serum Albumine solution.  Individual channels contain a $4.5\times 4.5\,$mm$^2$ region where $2R=25\,\mu$m diameter circular pillars are arranged in either hexagonal or slightly randomised square lattices of spacing $75\,\mu$m. 
Identical results were obtained with the two geometries.
Cells were imaged under either brightfield or phase contrast illumination with a Nikon TE2000-U inverted microscope fitted with a long-pass filter (cutoff wavelength $765\,$nm, Knight Optical, UK) to prevent phototactic stimulation. High throughput measurements of CR scattering off individual pillars are based on low magnification ($10\times$, Ph1, NA 0.25) low frame rate ($50\,$fps) recordings (Pike F-100B, AVT). 
Cells' trajectories during the $>80$k scattering events recorded for CC125 ($>45$k for SHF1) were digitised with standard particle tracking routines \cite{tracking, kilfoil}.
Approximately $300$ high magnification ($40\times$, NA 1.3), high frame rate (1200 fps, Phantom V 5.2, Vision Research) movies complemented previous measurements.
Experimentally, it is necessary to adopt a suitable criterion to define the beginning and end of the scattering. This needs to balance fully capturing the scattering process with reducing the impact of intrinsic swimming noise. We choose to define the scattering with reference to a circular impact area, concentric to the post and extending from its surface a distance $\sim3\,\mu$m larger than the flagellar length, giving a radius of $27\,\mu$m  for CC125 (SHF1: $25\,\mu$m) (see Fig.~\ref{fig1}b).
For each trajectory, the incoming and outgoing angles $(\thetain,\thetaout)$, are defined as the signed angle between the local radial direction and the incoming/outgoing swimming directions. The latter are calculated by a linear fit to the 5 trajectory points immediately external to the impact area. 
The total deflection angle is indicated with $\beta$ (sign conventions: Fig.~\ref{fig1}b). 

Fig.~\ref{fig1}c shows the experimentally determined conditional probabilities $p(\thetaout|\thetain)$. We will focus on the average $\avgthout(\thetain) = \int \thetaout\,p(\thetaout|\thetain)d\thetaout$. For both strains, this function is well described by 
\begin{equation}
 \avgthout(\thetain) =
  \begin{cases}
   \theta^*_{\textrm{out}} & \text{if } \thetain < \theta^*_{\textrm{in}} \\
   m\thetain+q       & \text{if } \thetain \geq \theta^*_{\textrm{in}}
  \end{cases}
\label{eq:thetaout}
\end{equation}
with the parameters summarised in Table~\ref{tab:values}.
Qualitatively similar results are borne out by the simulations of a three beads puller swimmer, our reference minimal model for CR, which includes hydrodynamics (see Table~\ref{tab:values}. Details in \cite{SupplMat}).
We will refer to $\thetain>\theta^*_{\textrm{in}}$ and $\thetain<\theta^*_{\textrm{in}}$ as the hydrodynamic and contact regimes respectively, although it will be seen that hydrodynamic forces play a  role also for $\thetain<\theta^*_{\textrm{in}}$. 
It should be kept in mind that Eq.~\ref{eq:thetaout} ignores the presence of a transition region between these regimes, which for CC125 happens over a $\sim20^\circ$ wide range of $\thetain$ (SHF1: $\sim15^\circ$). Within this region, for both strains, $\avgthout(\thetain)$ deviates from Eq.~\ref{eq:thetaout} by a small margin ($\leq3.4^\circ$, i.e. an error of $\lesssim6\%$). 
These small errors justify our coarsened but conceptually convenient approach.
\begin{figure}[b]
\includegraphics[width = 0.9\columnwidth ]{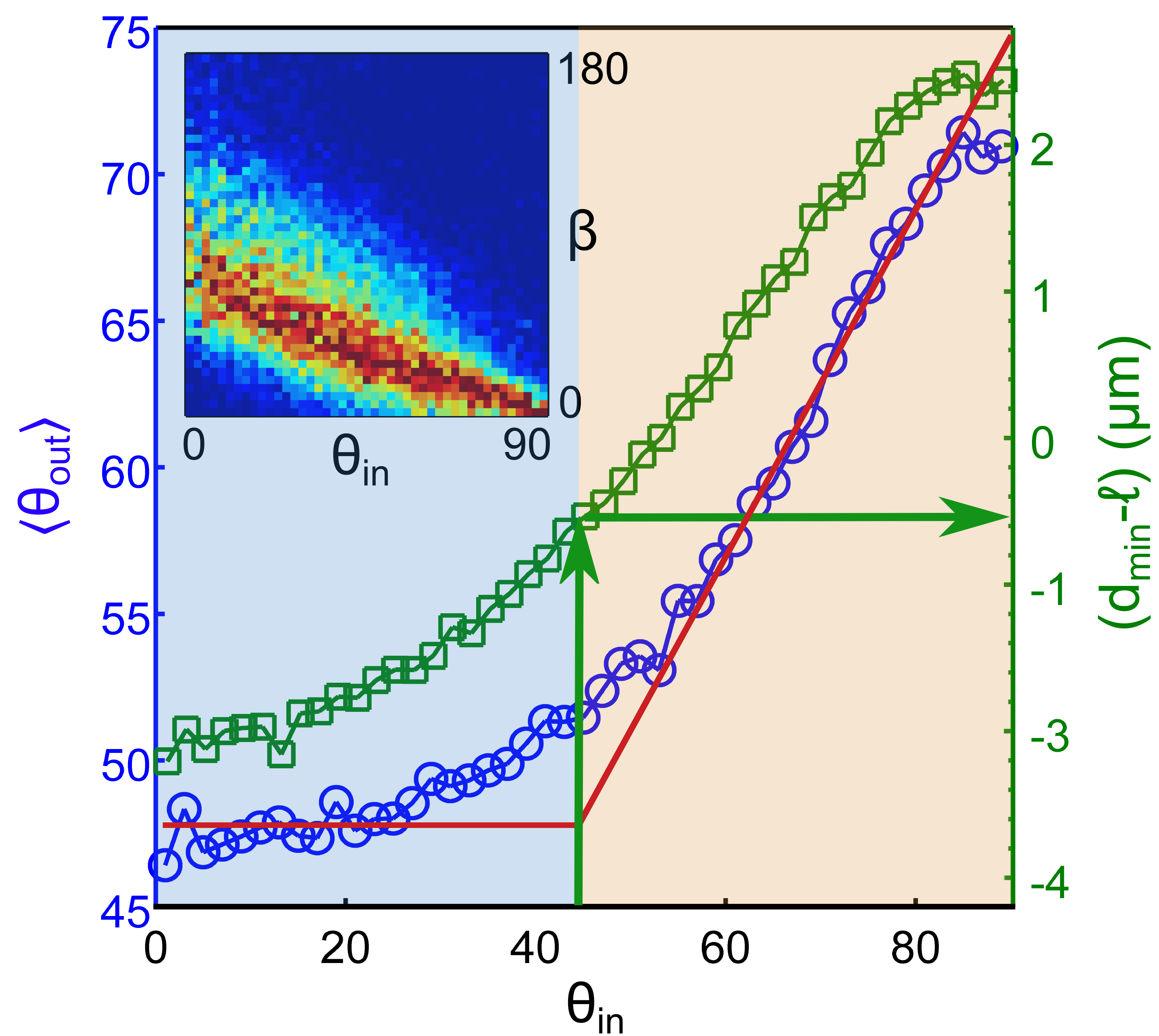}
\caption{Contact and hydrodynamic scattering (CC125). $\avgthout$ (blue circles), and $(\textrm{d}_{min}-\ell)$ (green squares) vs. $\thetain$. $\thetain=\theta^*_{\textrm{in}}$ corresponds to $\ell-\textrm{d}_{min}\simeq0.5\,\mu$m. Solid red line: Eq.~\ref{eq:thetaout}. (Inset) Deflection angle $\beta$ vs. $\thetain$. SHF1: see \cite{SupplMat}.}
\label{fig2}
\end{figure}

\begin{figure}[t]
\includegraphics[width = 0.95\columnwidth ]{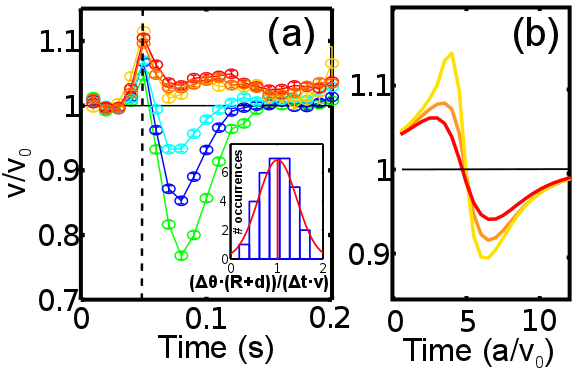}
\caption{Velocity during scattering $v(t)$ (CC125). (a) $v(t)/v_0$ for: $\thetain = 0-10 \degree $  ($\color{green}\circ$), $10-20 \degree$ ($\color{blue}\circ$), $20-30 \degree$ ($\color{cyan}\circ$), $50-60 \degree$ ($\color{yellow}\circ$), $60-70 \degree$ ($\color{amber}\circ$), $70-80 \degree$ ($\color{red}\circ$). Dashed line: scattering corona boundary. Inset: Experimental distribution of $\Delta\theta\,(R+d)/(\Delta t\,v)$ during recovery stage ($\thetain\in[0,35^{\circ}]$). Red line: Gaussian fit (mean: $1.02\pm0.07$; st. dev.: 0.42). (b) Hydrodynamic regime simulations: $v(t)/v_0$ with hydrodynamics ($\color{yellow}-$ $62^{\circ}$; $\color{amber}-$ $68^{\circ}$; $\color{red}-$ $71^{\circ}$) and without ($-$). $a$: swimmer radius.}
\label{fig3}
\end{figure}

Within the hydrodynamic regime, $\avgthout$ depends linearly on $\thetain$ with a slope $m\simeq 0.6$ (see Table~\ref{tab:values}).  
Qualitatively, $m\neq 1$ signals an interaction. Theoretical studies based on far-field hydrodynamics for puller microswimmers skimming off planar and spherical surfaces \cite{lauga09,spagnolie12, ardekani14, spagnolie15}, predict consistently a repulsive reorientation of the microorganism's trajectory.  Indeed, this can be directly observed from the angular deflection $\beta(\thetain)$ measured in our experiments (Fig.~\ref{fig2} inset).
$\beta$ ranges from $(33.5\pm 0.87)^\circ$ for $\thetain=57^\circ$ (just past the transition) down to $(5.1\pm1)^\circ$ for $\thetain =84^\circ$ (CC125) (SHF1: from $(42.6\pm 1)^\circ$ to $(6.9\pm1.3)^\circ$).
For the same range of $\thetain$, the minimal swimmer separation from the pillar surface, $\textrm{d}_{min}$, increases from $11.53\pm 0.04\,\mu$m to $13.65\pm 0.07\,\mu$m (CC125; Fig.~\ref{fig2}) (SHF1: from $9.42\pm 0.04\,\mu$m to $10.73\pm 0.03\,\mu$m).
Within most of the hydrodynamic regime, then, $\textrm{d}_{min}$ is larger than the average flagellar length $\ell = 11.2\pm 0.2\,\mu$m (CC125) (SHF1: $\ell = 9.36\pm 0.22\,\mu$m). Consequently, the observed interaction can only be ascribed to hydrodynamic forces (from which the name ``hydrodynamic'' regime). 
These cause also an increase of the swimming speed $v$ by up to $\sim10\%$ with respect to the average speed far from pillars, $v_0$, as the cell enters the corona (Fig.~\ref{fig3}a). This effect can be clearly seen in our simulations, but only when hydrodynamics is present (Fig.~\ref{fig3}b).
Conceptually similar evidence for the role of hydrodynamics has been reported for bacteria swimming at distances from a planar wall larger than the compound body and flagellar lengths \cite{molaei14}. We present it here for the first time for a much larger, puller-type eukaryotic microorganism.
Notice that, contrary to experiments, the simulations also show the swimmer decelerating as it leaves the pillar. The difference suggests that flagella increase their power output under a moderate load increase, as observed already for CR in \cite{minoura95}.

As $\thetain$ decreases, so does $\textrm{d}_{min}$ (Fig.~\ref{fig2}), and direct flagellar contact becomes important. The transition from hydrodynamic to contact regimes, then, is at a critical angle $\thetain=\theta^*_{\textrm{in}}$ where hydrodynamic and steric contributions to the deviation are equivalent. In our experiments, this happens when $\ell-\textrm{d}_{min}\simeq 0.4\,\mu$m ($(\ell,\textrm{d}_{min})=(11.2,10.7)\,\mu$m for CC125; $(9.36,9)\mu$m for SHF1). This is strikingly similar to the length of the flagellar tip ($\sim0.5\,\mu$m), where outer microtubule doublets progressively disappear leaving only the central pair \cite{harris09}. The tip is then likely significantly softer than the standard axoneme and therefore unable to provide a force sufficient to dominate the interaction with the wall.

Within the contact regime ($\thetain<\theta^*_{\textrm{in}}$),  $\textrm{d}_{min}<\ell$ and the alga makes close contact with the pillar. Following the average profile of the velocity $v(t)$ during the scattering (Fig.~\ref{fig3}a), the process can be divided in three stages. The first corresponds to the initial collision of the microalga with the post, which appears as a sudden deceleration  lasting between $40\,$ms ($\thetain\in [20^\circ, 30^\circ]$) and $60\,$ms ($\thetain\in [0^\circ, 10^\circ]$). Steric arguments \cite{elgeti13} would predict $v_0\sin(\thetain)$ (angle in radians) as the minimal speed. We observe a decrease of significantly  smaller magnitude (Fig.~\ref{fig3}a), reaching at most $\sim 25\%$ for almost head-on events. The discrepancy is due to partial cell reorientation during slowing down, possibly due to a combination of hydrodynamics and direct flagellar-wall contact. The reorientation is completed during the second stage, ending with a fully recovered speed ($v(t)= v_0$) and the cell aligned parallel to the pillar surface. Steric interactions imply that during this recovery, the angle $\theta$ that the cell makes with the local surface normal should obey (angles in radians) $\dot{\theta} = v(\theta)/(R+d)$, where $d$ is the swimmer-surface separation. Assuming $v(\theta) = v_0\sin(\theta)$, the angular variations within this stage, $\{\Delta\theta_j\}$, can be obtained from the instantaneous speed $\{v_j\}$. These should then satisfy $\left(\textrm{asin}(v_{j+1}/v_0)-\textrm{asin}(v_{j}/v_0)\right) = \Delta t(v_{j+1}+v_{j})/2(R+d)$, where $\Delta t = 20\,$ms is the inverse frame rate. Fig.~\ref{fig3}a inset shows that this is the case, supporting the interpretation that cells reorient by simply sliding over the convex, curved surface.
At the same time, the high speed movies \cite{SupplMat} reveal that at the end of the recovery stage the cell's flagellar plane is consistently parallel to the pillar surface. This subtle detail, probably resulting from direct contact interactions, is important in the final scattering stage, as we will now discuss.
Having completed its reorientation, the alga could be expected to simply swim away at speed $v_0$. 
For $\thetaout\sim50^\circ$ and starting at $\textrm{d}_{min}\simeq 9\,\mu$m from the pillar surface,  CC125 would take $\sim90\,$ms to exit the corona. Experimentally, instead, this stage lasts substantially longer: $170\pm8\,$ms.  
Within this time, the alga -spinning at a frequency of $1.78\pm\,0.4$Hz- can complete slightly more than 1/4 turn around its axis. 
Given the initial orientational bias, the flagellar plane should now be perpendicular to the wall.
Indeed, the high speed movies show clearly that cells leave the pillar with their flagellar plane always perpendicular to the surface.
This configuration maximises direct flagellar interaction with the obstacle, leading to a $\avgthout$ selected by the simple geometrical rule proposed in \cite{kantsler13}. For a measured body radius $a = 5.7\pm 0.1\,\mu$m, this would predict $\avgthout\simeq 90^\circ - \textrm{arctan}\left(\ell/2\,a\right) = 45.51^\circ$ (CC125), which compares very well with the experimental value $\theta^*_{\textrm{out}}=48^\circ$ (SHF1: $50.61^\circ$ vs. $50.76^\circ$).  
The duration of this stage, remarkably constant throughout the contact regime ($<5\%$ variation), can then be understood as the total time required for the cell to spin by $90^\circ$ ($\sim125\,$ms) and then swim away at the observed angle $\theta^*_{\textrm{out}}$ ($\sim50\,$ms). 
The simulations confirm the effect of the body rotation in helping an entrapped swimmer scatter off when captured at the pillar surface (see supporting movie).

This process hinges on a mechanism bending CR trajectories towards the pillar, an effective attraction eventually opposed by direct flagellar contact with the obstacle.
A quantitative measure of such interaction is given by the radius of curvature of the experimental trajectories, $\rho_{exp}$. From the evolution of the swimmer's distance to the pillar centre during the last stage of the scattering, we obtain $\rho_{exp} = 46\pm 8\,\mu$m \cite{SupplMat}.
At the same time, approximating CR's body as a sphere of radius $a$, near-field hydrodynamic torques (HT) can be estimated using lubrication theory \cite{KimandKarrilla} to give a radius of curvature
\begin{equation}
\rho_{\textrm{HT}} = a\left( \frac{10(1+\delta)^2}{\delta(4+\delta)} \right)\frac{1}{\ln(1/\epsilon)}
\label{radius_lubrication}
\end{equation}
where $\delta$ and $\epsilon$ are pillar radius and the gap between the swimmer and the pillar surface, non-dimensionalised by $a$.
For the experimentally determined values $(\delta= 2, \epsilon= 0.5)$ Eq.~\ref{radius_lubrication} gives $ \rho_{\textrm{HT}}= 65\,\mu$m.
This simple estimate reveals that lubrication forces alone already provide $\sim70\%$ of the observed effective torque, supporting the interpretation of a fundamentally hydrodynamic origin for the observed effective attraction. Extra torques come possibly from a combination of further hydrodynamic contributions beyond lubrication, and unequal performance of the two flagella. 
Although this suggests that sufficiently large pillars should trap CRs hydrodynamically, in the present case the scattering is terminated by contact forces as soon as the flagellar plane becomes perpendicular to the surface. 
 As a result, we predict that scattering events in the contact regime should have the same duration even for larger obstacles. This is supported by experiments with $40\,\mu$m-radius pillars \cite{SupplMat}, and it is compatible with previous results from scattering off a plane \cite{kantsler13}. Our observations, then, differ from theoretical estimates of microorganismal capture by curved obstacles, based either on hydrodynamic \cite{spagnolie15, schaar15} or steric \cite{elgeti13} interactions, where escape results purely from noise (although \cite{schaar15} discusses using a constant torque to mimick flagellar activity). 
\begin{figure}[t]
\includegraphics[width = 0.95\columnwidth ]{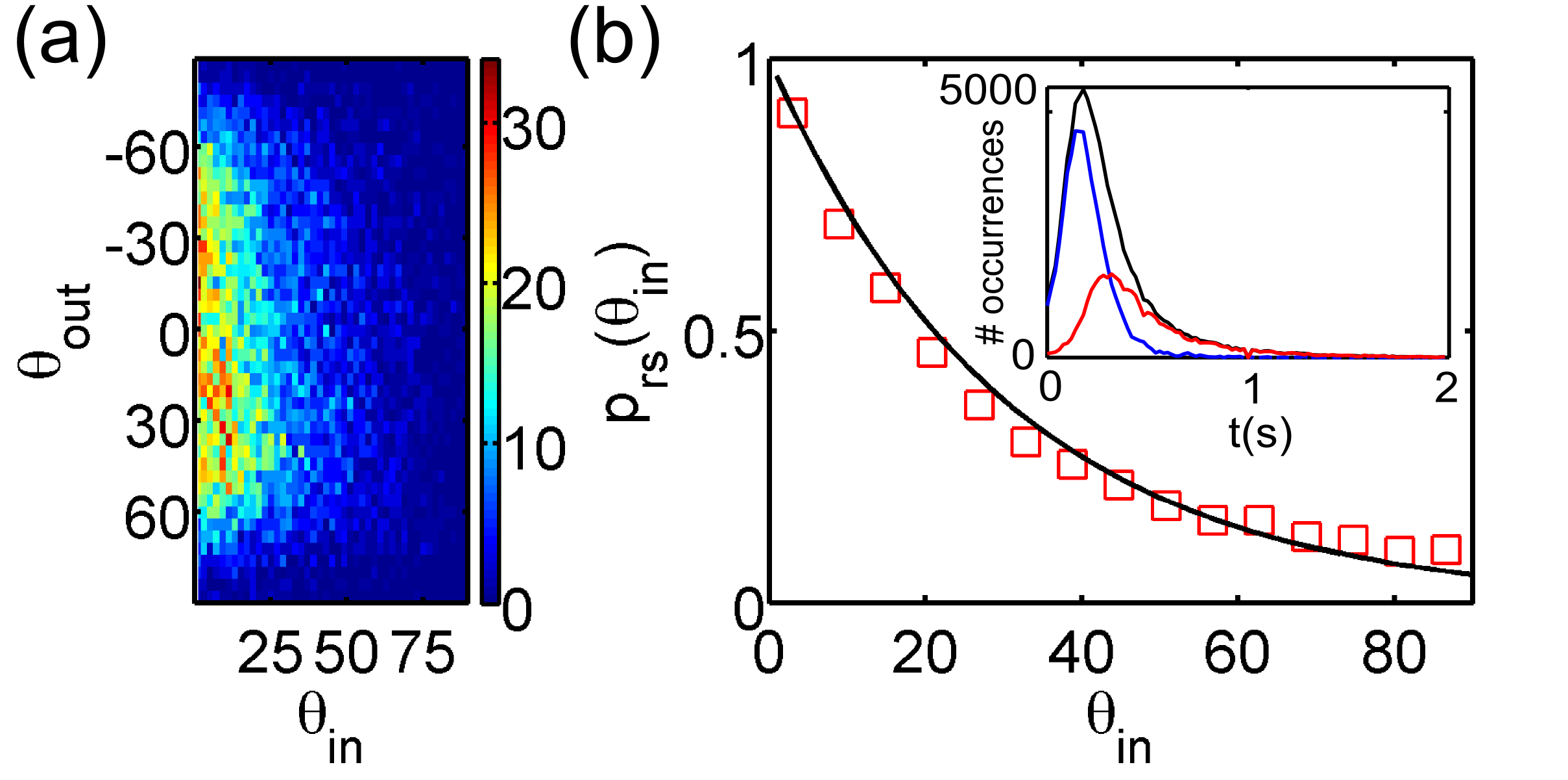}
\caption{Random scattering (CC125). (a) Distribution of $\thetaout$ vs. $\thetain$ for random scatterings. (b) $P_{\textrm{rs}}(\thetain)$ vs. $\thetain$ experiments ($\color{red}\square$) and fit ($\color{black}-$). (Inset) Distribution of all ($\color{black}-$), and random ($\color{red}-$) scattering events' duration. The difference represents deterministic events ($\color{blue}-$).}
\label{fig4}
\end{figure}

The scattering discussed so far proceeds according to a largely predictable dynamics and could be described as deterministic.
Together with this, however, we observe a qualitatively different type of interaction, which we call random scattering. Random scatterings are characterised by a prolonged ($\sim 500\,$ms) almost head-on collision of the alga with the obstacle, resulting in multiple events of direct flagellar interaction with the surrounding surfaces, often including the upper and lower boundaries of the microfluidic chamber. A long duration is in fact the hallmark of random scatterings, and was used to distinguish these from deterministic events \cite{SupplMat}. 
The outcome of such intrinsically complex dynamics is simple: $\thetaout$ is uniformly distributed across the available range of values, independently of $\thetain$ (Fig.~\ref{fig4}a), a behaviour in sharp contrast to that of minimal models of puller microorganisms \cite{spagnolie15, schaar15}. 
At the same time, the probability $P_{\textrm{rs}}$ of performing random rather than deterministic scattering, depends strongly on $\thetain$. Empirically, we find $P_{\textrm{rs}}(\thetain)\propto \textrm{exp}(-\thetain/\thetars)$, where $\thetars=33.3^\circ$ for CC125 (Fig.~\ref{fig4}b, and Table~\ref{tab:values}). This distribution can be recovered within a simple model where the initial dynamics of the orientation angle $\theta$ follows an advection-diffusion process in $[0^{\circ},90^{\circ}]$ with absorbing boundaries leading to either random ($0^{\circ}$) or deterministic ($90^{\circ}$) scattering.
Then $\thetars = D_r/\omega$, where  $D_r$ and $\omega$ are the (rotational) diffusion and drift respectively. 
From the deterministic scattering, we can estimate the effective drift towards $90^{\circ}$ as $\omega = \langle\dot{\theta}\rangle=2v_0/\pi\,R \simeq 5\,$rad/s. The experimental $\thetars$ then implies $D_r\simeq3\,$rad$^2$/s (CC125), which agrees well with $D_r\simeq 2\,$rad$^2$/s previously measured for the related species {\it Chlamydomonas nivalis} \cite{hill97}. Assuming the same $D_r$ for CC125 and SHF1, the model predicts also that $\theta_{\textrm{rs}}^{\textrm{CC125}}/\theta_{\textrm{rs}}^{\textrm{SHF1}}$$(\simeq0.67)$ should equal $v_0^{\textrm{SHF1}}/v_0^{\textrm{CC125}}$$(\simeq0.615)$, indeed verified experimentally within $\lesssim 10\%$. 

We have presented the first experimental study of the interaction of the model microalga {\it Chlamydomonas reinhardtii}, commonly regarded as a prototypical puller microswimmer, with circular obstacles. The experiments reveal that the scattering is not simply steric, as previously suggested \cite{kantsler13}, but follows qualitatively different rules depending on the angle of incidence, with direct evidence of purely hydrodynamic interaction at large angles, and a multi-stage steric/hydrodynamic process at small angles. The latter is terminated by direct flagellar contact with the surface preventing the extended trapping around the convex structure predicted by minimal models \cite{elgeti13, spagnolie15, schaar15}, a behaviour recapitulated by our simulations. The ability to avoid long-term trapping independently of obstacle size and shape might in fact represent a significant advantage for a soil alga like {\it Chlamydomonas}, which in nature needs to navigate a heterogeneous porous material. Together with these deterministic interactions, we report the existence of random scatterings, and propose a mechanism to explain their likelihood. 
Although these events could be specific to our experimental configuration, their existence still suggests that front-mounted flagella  pose a challenge to coarse grained descriptions of CR-like microorganisms' interactions with surfaces.

\end{document}